\documentclass[12pt,letterpaper]{article}
\usepackage{epsfig} 
\usepackage{graphicx}
\usepackage{graphics}  
\usepackage{cite}      % bibliographic sitations 
\usepackage{floatflt}  
\usepackage{fancybox}  %allows for shadow boxes etc.
\usepackage{color}     %allows use of colors (RBG only, I think)

\begin{document}
%
%%%%%%%%%%%%%%%%%%%%%%%
% Define new commands %
%%%%%%%%%%%%%%%%%%%%%%%
%
\newenvironment{myitemize}[2]
{ \begin{small}
  {\large\tt\bf #1} { \sc #2 } \vspace{-.2cm}
  \begin{itemize}
  \setlength{\itemsep}{-4pt}
  \setlength{\parsep}{0pt}
  \setlength{\topsep}{0pt}
  \setlength{\partopsep}{0pt}
}
{\end{itemize}\end{small}}
\newcommand{\myitem}[3]{ \item {\footnotesize\tt #1}~{\bf\tt #2}: {\it #3} }

\newcommand{\MEG}{MEG}
\newcommand{\SHG}{SHG}

%
%%%%%%%%%%%%%%%%%%%%%%%%%%%%%%%%%%%%%%%%%%%%%%%%%%%%%%%%%%%%%%%%%%%%%
%
\pagenumbering{arabic}

\begin{center}

\vspace{1cm}

{\Large\bf Generic User Process Interface for Event Generators}

\vspace{1cm}

{ E.\ Boos, M.\ Dobbs, W.\ Giele,
I.\ Hinchliffe, J.\ Huston, V.\ Ilyin, \\
J.\ Kanzaki, K.\ Kato, Y.\ Kurihara, 
L.\ L\"onnblad, M.\ Mangano, S.\ Mrenna, \\
F.\ Paige, E.\ Richter-Was,
M.\ Seymour, T.\ Sj\"ostrand, B.\ Webber, D.\ Zeppenfeld }
% \footnote{
%  If your name is missing and you would like it here, please say so. \\
%  If your name is here and you would like it missing, please say so. \\
%  Please verify the references---make sure your MC is properly referenced. }

\vspace{.5cm}

September 7, 2001

\vspace{.5cm}

\end{center}
\begin{abstract}
Generic Fortran common blocks are presented for use by High Energy
Physics event generators for the transfer of event configurations from
parton level generators to showering and hadronization event
generators.
\end{abstract}

Modularization of High Energy Particle Physics event generation is
becoming increasingly useful as the complexity of Monte Carlo programs
grows. To accommodate this trend, several authors of popular Monte
Carlo and matrix element 
programs attending the {\it Physics at TeV Colliders Workshop}
in Les Houches, 2001 have agreed on a generic format for the transfer
of parton level event configurations from matrix element event generators 
(\MEG) to showering and hadronization event generators (\SHG).

\begin{center} \color{blue}
\begin{tabular}{rcl}
  CompHEP\cite{Pukhov:1999gg}     & &        \\
  Grace\cite{Ishikawa:1993qr}     & & Herwig\cite{Corcella:2001bw} \\
  MadGraph\cite{Stelzer:1994ta}   & \hspace{1cm} $\Rightarrow$
  				  \hspace{1cm} & Isajet\cite{Baer:1999sp} \\
%%%       SusyGen\cite{Katsanevas:1998fb} 
  VecBos\cite{Berends:1991ax}     & & Pythia\cite{Sjostrand:2001wi} \\
  WbbGen\cite{Caravaglios:1999yr} & & \ldots \\
  \ldots    & &        \\
\end{tabular}
\color{black} \end{center}

Events generated this way are customarily called user (or
user-defined) processes, to distinguish them from the internal
processes that come with the \SHG. Specific solutions are already in
use, including an interface of WbbGen with
Herwig~\cite{Mangano:2001xp} and an interface of CompHEP with
Pythia~\cite{Belyaev:2000wn}---that experience is exploited here.

%% I know of atleast two others:   VecBos --> Herwig (see joey's mail)
%%                             and Grace --> Pythia  (see spires)

An implementation of the user process interface has been included in
Pythia~6.2, described (with an example) in Ref.~\cite{Sjostrand:2001doc}.

The user process interface discussed here is not intended as a
replacement for \mbox{ {\tt HEPEVT}}~\cite{HEPEVT}, 
which is the standard Fortran
common block for interface between generators and analysis/detector
simulation. The user process common blocks address the communication
between two event generators only, a \MEG\ one and a \SHG\ one, and not the 
communication of event generators with the outside world. 

In the course of a normal event generation run, this communication occurs 
at two stages: (1) At initilization, to establish the basic parameters 
of the run as a whole. (2) For each new event that is to be 
transferred from the \MEG\ to the \SHG. Each of these two stages here
corresponds to its own Fortran common block.\footnote{ An interface in 
C++ has been developed in Ref.~\cite{Dobbs:2001ck} and contains similar 
information content as that discussed here.} These common blocks are 
described in detail in the next two sections, followed by some examples.

One can also foresee that each stage will be associated with its own
subroutine, called from the \SHG, where information is put in the
respective common block, based on output from the \MEG. The details of
these subroutines are likely to be specific to a given \MEG\ and may
also be specific to a given \SHG.  The subroutine names {\tt UPINIT} and
{\tt UPEVNT} (each with no arguments) were chosen for the
Pythia~6.2 implementation.  They are intended to
be generic (the usual {\tt PY} prefixes are omitted), and only dummy
versions are packaged with the program.  It is recommended that other
\SHG\ authors use the same dummy routine names (with zero arguments)
such that for simple cases which do not require intervention 
`by hand', \MEG\ authors will be able to interface several \SHG s with a
single set of routines. Example routines are presented in the Pythia
documentation~\cite{Sjostrand:2001doc}.

In general, a user process run may consist of a number of subprocesses,
each denoted by a unique integer identifier. If the user
wishes to have the \SHG\ unweight events using acceptance-rejection and/or
mix together events from different processes, then the user process
author will need to supply a subroutine that is able to return an event of
the requested subprocess type to the \SHG\ on demand. 
The author may choose to organize the subroutine to
generate the event `on the fly', or to read the event from a file
stream (with a separate file stream for each subprocess).
The \SHG\ will also need information about the subprocess cross section
and/or maximum event weight to select which process is generated next
and for acceptance-rejection. This information will need to be
known from the onset (and could, for example, be determined in advance
from an initialization run). Alternatively, the user may already have 
a proper mixture of subprocesses from the \MEG\ and only wish the \SHG\
to process events in the order they are handed in. We therefore allow for 
several different event weight models.

If extra information is needed for a specific user implementation,
then a implementation-specific common block should be
created. The meaning of the user process common block
elements should not be overloaded, as this would defeat the generic
purpose.

The descriptions in this paper are intended for event generator
authors and may appear complex---most of the details will be
transparent to the casual user.

\section{`User Process' Run Information}

        The run common block contains information which pertains to a
        collection of events. \\
        In general this information is process dependent and it is
        impossible to include everything in a generic common
        block. Instead only the most general information is included
        here, and it is expected that users will have to intervene
        `by hand' for many cases (i.e.\ a user may need to specify
        which cutoffs are used to avoid singularities, which jet 
        clustering algorithm has been used to recombine partons in a
        next-to-leading-order calculation, the effective parton masses, 
        \ldots).

\begin{verbatim}
      integer     MAXPUP
      parameter ( MAXPUP=100 )
      integer     IDBMUP, PDFGUP, PDFSUP, IDWTUP, NPRUP, LPRUP
      double precision EBMUP, XSECUP, XERRUP, XMAXUP
      common /HEPRUP/ IDBMUP(2), EBMUP(2), PDFGUP(2), PDFSUP(2),
     +                IDWTUP, NPRUP, XSECUP(MAXPUP), XERRUP(MAXPUP), 
     +                XMAXUP(MAXPUP), LPRUP(MAXPUP)
\end{verbatim}
%      CHARACTER*28 CHPRUP
%      COMMON/HEPCUP/CHPRUP(MAXPUP)

\vspace{.5cm}
        
\begin{myitemize}{HEPRUP}{ `User Process' Run Common Block }

   \myitem{parameter} { MAXPUP=100 }
        {maximum number of different processes to be interfaced at one
        time} 

\hspace{-.5cm} {\bf \underline{ Beam Information } } \\
        Beam particle 1 (2) is defined as traveling along the +Z (--Z)
        direction.
   \myitem{integer} { IDBMUP(2) }
        {ID of beam particle 1 and 2 according to the Particle Data
        Group convention~\cite{Garren:2000st}}
   \myitem{double} { EBMUP(2) }
        {energy in GeV of beam particles 1 and 2}
   \myitem{integer} { PDFGUP(2) }
        { the author group for beam 1 and 2, according to the Cernlib
        PDFlib~\cite{Plothow-Besch:1993qj} specification }
   \myitem{integer} { PDFSUP(2) }
        { the PDF set ID for beam 1 and 2, according to the Cernlib
        PDFlib specification } \\
    For $e^+e^-$ or when the \SHG\ defaults are
        to be used, PDFGUP=--1, PDFSUP=--1 should be specified. \\
    The PDFlib enumeration of PDFs is sometimes out of date, but it is the
        only unique integer labels for PDFs available. In the case
        where a PDF not included in PDFlib is being used, this
        information will have to be passed `by hand'.

\hspace{-.5cm} {\bf \underline{ Process Information } }

   \myitem{integer} { IDWTUP }
	{master switch dictating how the event weights
        (XWGTUP) are interpreted} \\
        The user is expected to pick the most appropriate event weight 
        model for a run,
        given the \MEG\ input at hand and the desired output. Normally 
        the \SHG\ should be able to handle all of the models.\\   
	A summary of the IDWTUP switches is presented in
	Table~\ref{t_IDWTUP}.
        \begin{itemize}
        \setlength{\itemsep}{-4pt} \setlength{\parsep}{0pt}
        \setlength{\topsep}{0pt} \setlength{\partopsep}{0pt} 

        \item[+1] Events are weighted on input and the \SHG\ is asked
        to produce events with weight +1 as output. XSECUP and XERRUP
        need not be provided, but are calculated by the \SHG. XWGTUP
        is a dimensional quantity, in pb, with a mean value converging
        to the cross section of the process.  The \SHG\ selects the
        next subprocess type to be generated, based on the relative
        size of the XMAXUP(i) values. The user-supplied interface routine
        must return an event of the requested type on demand from the
        \SHG, and the maximum weight XMAXUP (or a reasonable
        approximation to it) must be known from the onset.
	A given event is accepted with a probability
        XWGTUP/XMAXUP(i). In case of rejection, a new event type and a
        new event are selected.  If XMAXUP(i) is chosen too low, such
        that XWGTUP violates the XMAXUP(i), the \SHG\ will issue a
        warning and update XMAXUP(i) with the new maximum weight
        value.
	If events of some types are already available unweighted, then
        a correct mixing of these processes is ensured by putting
        XWGTUP = XMAXUP(i). In this option also the internal \SHG\
        processes are available, and can be mixed with the external
        ones.  All weights are positive definite. $k$-factors may be
        included on an event by event basis by the user process 
        interface by re-scaling the XWGTUP for each event.

        \item[--1] Same as above (IDWTUP=+1), but the event weights
	may be either positive or negative on input, and the \SHG\
	will produce events with weight +1 or --1 as 
        output.\footnote{Negative-weight events may occur e.g.\ in 
        next-to-leading-order calculations. They should cancel against
        positive-weight events in physical distributions. The details
        of this cancellation are rather subtle when considered in
        the context of showers and hadronization, however, and a
        proper treatment would require more information than discussed 
        here. The negative-weight options should therefore be used with 
        some caution, and the negative-weight events should be a 
        reasonably small fraction of the total event sample.}
        A given event would be accepted with a probability
	$|$XWGTUP$|$/$|$XMAXUP(i)$|$ and assigned weight {\tt
	sign}(1,XWGTUP), where the {\tt sign} function transfers the
	sign of XWGTUP onto 1. A
	physics process with alternating cross section sign must
	be split in two IDPRUP types,\footnote{The motivation for this
	requirement is best understood with a simple example: imagine
	two subprocesses with the same cross section. The first
	process includes events with both positive and negative event
	weights such that two events out of three have weight +1 and the
	third --1. All events from the second process have positive
	weight +1. In this scenario these two processes should be
	`mixed' with proportions 3:1 to account for the cancellations
	that occur for the first process. The proportions for the
	mixing are communicated to the \SHG\ by supplying the positive
	and negative contributions to the cross section separately.
	}
	based on the sign of
	the cross section, such that all events of a particular IDPRUP
	have the same event weight sign.  Also the XMAXUP(i) values must
	be available for these two IDPRUP types separately, so that
	$|$XMAXUP(i)$|$ gives the relative mixing of event types, with
	event acceptance based on $|$XWGTUP$|$/$|$XMAXUP(i)$|$.

	\item[+2] Events are weighted on input and the \SHG\ is asked
        to produce events with weight +1 as output. The \SHG\ selects
        the next subprocess type to be generated, based on the
        relative size of the XSECUP(i) values.  The user-supplied
        interface routine must return an event of the requested type
        on demand from the \SHG. The cross sections XSECUP(i) must be
        known from the onset.
%though it may be poorly known at the beginning and gradually 
%updated by the user process
%	This would imply the presence of a separate file of events,
%        alternatively a separate generation algorithm, for each
%        process type.  
	A given event is accepted with a probability
        XWGTUP/XMAXUP(i).  In case of rejection, a new event of
        the same type would be requested.  In this scenario only the
        ratio XWGTUP/XMAXUP(i) is of significance. If events of
        some types are already available unweighted, then a correct
        mixing of these processes is ensured by putting XWGTUP =
        XMAXUP(i). A $k$-factor can be applied
        to each process by re-scaling the respective XSECUP(i)
        value at the beginning of the run, but cannot be given
        individually for each event. In this option also the internal
        \SHG\ processes are available, and can be mixed with the user
        processes.

        \item[--2] Same as above (IDWTUP=+2), but the event weights
	may be either positive or negative on input, and the \SHG\
	will produce events with weight +1 or --1 as output.  A
	physics process with alternating cross section sign must
	therefore be split in two IDPRUP types, based on the sign of
	the cross section, such that all events of a particular IDPRUP
	have the same event weight sign.  Also the XSECUP(i) and
	XMAXUP(i) values must be available for these two IDPRUP types
	separately, so that $|$XSECUP(i)$|$ gives the relative mixing
	of event types, with event acceptance based on
	$|$XWGTUP$|$/$|$XMAXUP(i)$|$ and the total cross section of
	the two IDPRUP types combined given by XSECUP(i)+XSECUP(j).

        \item[+3] Events are unweighted on input such that all events
	come with unit weight XWGTUP=+1. The \SHG\ will only ask for
	the next event.  If any mixing or unweighting is desired, it
	will have to be performed by the user process interface.  The 
        \SHG\ will not reject any events (unless it encounters other 
        kinds of problems).  If a $k$-factor is desired, it is the
	responsibility of the user process interface. When events are  
        read sequentially from an already existing file, this would    
        imply one common $k$-factor for all processes. In this option it 
        is not possible to mix with internal \SHG\ processes.
 
        \item[--3] Same as above (IDWTUP=+3), but the event weights
	may be either +1 or --1 on input. A single process identifier
        (IDPRUP) may include events with both positive and negative
        event weights.

        \item[+4] Same as (IDWTUP=+3), but events are weighted
        on input and the average of the event weights (XWGTUP) is
        the cross section in pb. When histogramming results on analyzed
        events, these weights would have to be used.
        The \SHG\ will only ask for the next event
        and will not perform any mixing or unweighting. Neither XSECUP
        nor XMAXUP needs to be known or supplied. In this
	option it is not possible to mix with internal \SHG\
	processes.

        \item[--4] Same as (IDWTUP=+4), but event weights may be either
        positive or negative on input and the average of the event
        weights (XWGTUP) is the cross section. A single process identifier
        (IDPRUP) may include events with both positive and negative
        event weights.

        \end{itemize} 

\begin{table}[h]
\begin{center}
\begin{tabular}{c c c c c} \hline
        & event selection& control of        & XWGTUP &        \\
 IDWTUP & according to   & mixing/unweighting& input & output \\ \hline
+1      & XMAXUP         & \SHG\        & + weighted     & +1 \\
--1     & XMAXUP         & \SHG\        & $\pm$ weighted & $\pm$1 \\
+2      & XSECUP         & \SHG\        & + weighted     & +1 \\
--2     & XSECUP         & \SHG\        & $\pm$ weighted & $\pm$1 \\
+3      & ---            & user interface & +1             & +1 \\
--3     & ---            & user interface & $\pm$1         & $\pm$1 \\
+4      & ---            & user interface & + weighted     & + weighted \\
--4     & ---            & user interface & $\pm$ weighted & $\pm$ weighted \\
\hline
\end{tabular}
\caption{\label{t_IDWTUP} Summary of the options available for the
master weight switch IDWTUP.}
\end{center}
\end{table}

   \myitem{integer} { NPRUP }
        { the number of different user subprocesses } \\
	i.e.\ LPRUP and other arrays will have NPRUP entries, LPRUP(1:NPRUP)

   \myitem{double} { XSECUP(J) }
        { the cross section for process J in pb } \\
        This entry is mandatory for IDWTUP=$\pm$2.

   \myitem{double} { XERRUP(J) }
        { the statistical error associated with the cross section 
	of process J in pb }\\
	It is not expected that this information will be used by the
        \SHG, except perhaps for printouts.

   \myitem{double} { XMAXUP(J) }{ the maximum XWGTUP for process J } \\
	For the case of weighted events (IDWTUP=$\pm$1,$\pm$2), this entry is
        mandatory---though it need not be specified to a high degree
        of accuracy. If too small a number is specified, the \SHG\
	will issue a warning and increase XMAXUP(J).
        This entry has no meaning for IDWTUP=$\pm$3,$\pm$4. 

   \myitem{integer} { LPRUP(J) } {a listing of all user process IDs
        that can appear in IDPRUP of HEPEUP for this run} \\
	When
        communicating between the user process and \SHG, the LPRUP
        code will be used. Example: if LPRUP(1)=1022, then the \SHG\
        will ask for an event of type 1022, not 1.

%   \myitem{character*28} { CHPRUP(J) } 
%	{ character string of process name (optional) }
%	This entry may be used by the \SHG s for print statements.
%	This entry is contained in its own common block, HEPCUP.

\end{myitemize}

\begin{center}{\rule[1mm]{5in}{1mm}}\end{center} %%%%%line

\section{`User Process' Event Information}

\begin{verbatim}
      integer MAXNUP
      parameter ( MAXNUP=500 )
      integer NUP, IDPRUP, IDUP, ISTUP, MOTHUP, ICOLUP
      double precision XWGTUP, SCALUP, AQEDUP, AQCDUP, 
     +                 PUP, VTIMUP, SPINUP
      common /HEPEUP/ NUP, IDPRUP, XWGTUP, SCALUP, AQEDUP, AQCDUP,
     +                IDUP(MAXNUP), ISTUP(MAXNUP), MOTHUP(2,MAXNUP),
     +                ICOLUP(2,MAXNUP), PUP(5,MAXNUP), VTIMUP(MAXNUP),
     +                SPINUP(MAXNUP)
\end{verbatim}

\vspace{.5cm}

\begin{myitemize}{HEPEUP}{ `User Process' Event Common Block }

  \myitem{parameter} { MAXNUP=500 }
        { maximum number of particle entries }
%         Parton level processes will typically have only a few
%         entries. A relatively large number is specified to also allow for
%         the possibility of interfacing at the stage between the parton
%         shower and hadronization steps of event generation.

  \myitem{integer} { NUP } 
        {number of particle entries in this event} \\
	An event with NUP=0 denotes the case where the user process is unable
        to provide an event of the type requested by the \SHG\
	(i.e.\ if the user process
        is providing events to the \SHG\ by reading them sequentially
        from a file and the end of the file is reached).

  \myitem{integer} { IDPRUP } 
        {ID of the process for this event} \\
        The process ID's are not intended to be generic. The entry is
        a hook which the event generators can use to
        translate into their own scheme, or use in print statements
        (e.g.\ so that cross section information can be shown per
        process). \\
	When IDWTUP$=\pm1,\pm2$ the next process to be generated 
	is selected by the \SHG, and so IDPRUP is set by the \SHG. For 
	IDWTUP$=\pm3,\pm4$ the process is selected by the \MEG, and
        IDPRUP is set by the \MEG.

  \myitem{double} { XWGTUP }
        { event weight } \\
        \underline{weighted events}:
        if the user process supplies
        weighted events and the \SHG\ is asked to produce 
	unweighted events, this number
        will be compared against XMAXUP in the run common block HEPRUP
        for acceptance-rejection. \\
	\underline{unweighted events}:
        if the user process supplies events which have already been
        unweighted, this number should be set to +1 (-1 for negative
        weight events in e.g. a NLO calculation).

	The precise definition of XWGTUP depends on the master weight
        switch IDWTUP in the run common block. More information is
        given there.

  \myitem{double} { SCALUP } 
        {scale of the event in GeV, as used for calculation of  PDFs} \\
        If the scale has not been defined, this should be denoted
        by setting the scale to --1.

  \myitem{double} { AQEDUP } 
        {the QED coupling $\alpha_\mathrm{QED}$ used for this event 
        (e.g.\ $\frac{1}{128}$) }
        
  \myitem{double} { AQCDUP } 
        {the QCD coupling $\alpha_\mathrm{QCD}$ used for this event }

        When $\alpha_\mathrm{QED}$ and/or $\alpha_\mathrm{QCD}$ is not
        relevant for the process, or in the case where the user
        process prefers to let the \SHG\ use its defaults,
        AQEDUP=--1 and/or AQCDUP=--1 should be specified.
        
\hspace{-.5cm} {\bf \underline{ ID, Status, and Parent-Child History } }
  \myitem{integer} { IDUP(I) } 
        {particle ID according to Particle Data Group 
	convention~\cite{Garren:2000st}} \\
        undefined (and possibly non-physical) ``particles'' should be
        assigned IDUP=0 (i.e.\ the $WZ$ particle in the example given
        in the MOTHUP description below)

  \myitem{integer} { ISTUP(I) } {status code} 
        \begin{itemize}
        \setlength{\itemsep}{-4pt} \setlength{\parsep}{0pt}
        \setlength{\topsep}{0pt} \setlength{\partopsep}{0pt} 
        \item[--1] Incoming particle
        \item[+1] Outgoing final state particle
	\item[--2] Intermediate space-like propagator defining an 
	$x$ and $Q^2$ which should be preserved
        \item[+2] Intermediate resonance, Mass should be preserved
        \item[+3] Intermediate resonance, for documentation only\footnote{
		Treatment of ISTUP(I)=+3 entries may be generator 
		dependent (in particular see
       	 	Ref.~\cite{Sjostrand:2001doc} 
		for the special treatment in Pythia).}
        \item[--9] Incoming beam particles at time $t=-\infty$
        \end{itemize} 
        The recoil from a parton shower (including
        photon emission) needs to be absorbed by particles in the
        event. Without special instructions, this can alter the mass
        of intermediate particles. The ISTUP flag +2 allows the user
        process to specify which intermediate states should have their
        masses preserved, i.e.\ for $e^+e^- \rightarrow Z^0 h^0
        \rightarrow q \bar{q} b \bar{b}$, the $Z^0$ and $h^0$ 
        would be flagged with ISTUP=+2. \\
	The primary application of the ISTUP=--2 status code is deep
        inelastic scattering (a negative number is chosen for this
        status code because the propagator in some sense can be
        thought of as incoming). See the example below. \\
	The status code ISTUP=--9 specifying incoming beams is not
        needed in most cases because the beam particle energy and identity
        is contained in the HEPRUP run information common block. The
        primary application of this status code will be non-collinear
        beams and/or runs for which the beam energy varies event by
        event (note that using the --9 status code to vary the machine
        energy may produce problems if the \SHG\
	is asked to combine separate processes).  The
        use of ISTUP=--9 entries is optional, and is only necessary
        when the information in HEPRUP is insufficient. If entries
        with ISTUP=--9 are specified, this information will over-ride
        any information in HEPRUP.

  \myitem{integer} { MOTHUP(2,I) } 
        {index of first and last mother} \\
        For decays, particles will normally have only one mother. In
        this case either MOTHUP(2,I)=0 or MOTHUP(2,I)=MOTHUP(1,I).
        Daughters of a $2\rightarrow n$ process have 2 mothers.  This
        scheme does not limit the number of mothers, but in practice
        there will likely never be more than 2 mothers per particle.

        The history (intermediate particles) will be used by the
        \SHG s to decipher which
        combinations of particles should have their masses fixed and
        which particle decays should be ``dressed'' by the parton
        shower. Example: for $q\bar{q}'\rightarrow W^-Zg\rightarrow l^-
        \nu l^+ l^- g$, intermediate ``particles'' $WZ$, $W$, and $Z$
        could be specified with ISTUP=+2. Here the $WZ$ ``particle''
        would have its own entry in the common block with IDUP=0.  The
        showering generator would preserve the invariant masses of
        these ``particles'' when absorbing the recoil of the parton
        shower.

        In a case like $e^+e^- \rightarrow \mu^+\mu^-\gamma$
        proceeding via a
        $\gamma^*/Z^0$, where the matrix element contains an
        interference term between initial and final-state emission,
        this ambiguity in the parent-child history of the $\gamma$ has
        to be resolved explicitly by the user process.

  \hspace{-.5cm} {\bf \underline{ Color Flow } } \\
        A specific choice of color flow for a particular event is often
        unphysical, due to interference effects. However,
        \SHG s require a specific color state from which to begin
        the shower---it is the responsibility of the user process to
        provide a sensible choice for the color flow of a particular event.
        
  \myitem{integer} { ICOLUP(1,I) }
        {integer tag for the color flow line passing through the color of
        the particle}
  \myitem{integer} { ICOLUP(2,I) }
        {integer tag for the color flow line passing through the
        anti-color of the particle} \\
        The tags can be viewed as numbering the different color lines in
        the $N_C\rightarrow \infty$ limit.
        The color/anti-color of a particle are defined with respect to
        the physical time order of the process so as to allow a unique
        definition of color flow also through intermediate particles.\\
        This scheme is chosen because it has the fewest ambiguities, and
        when used with the history information, it supports Baryon
        number violation (an example is given below). \\
        To avoid confusion it is recommended that integer tags larger than 
        MAXNUP (i.e.\ 500) are used. The actual value of the tag has no meaning
        beyond distinguishing the lines in a given process.
        
  \hspace{-.5cm} {\bf \underline{ Momentum and Position } }
  \myitem{double} { PUP(5,I) }
        { lab frame momentum $(P_x, P_y, P_z, E, M)$ of particle in GeV } \\
        The mass is the `generated mass' for this particle, 
        $M^2=E^2-|\vec{p}|^2$ (i.e.\ not
        necessarily equal to the on-shell mass). The mass may 
        be negative, which denotes negative $M^2$ (i.e.\
        $M=2$ implies $M^2=4$ whereas $M=-2$ implies $M^2=-4$). \\
        Both $E$ and $M$ are needed for numerical reasons, the user
        should explicitly calculate and provide each one.

  \myitem{double} { VTIMUP(I) }
        { invariant lifetime $c\tau$ (distance from production to decay) in
        mm} \\
        Combined with the directional information from the momentum,
        this is enough to determine vertex locations.  Note that this
        gives the distance of travel for the particle from birth to
        death, in this particular event, and not its distance from the
        origin.

\hspace{-.5cm} {\bf \underline{ Spin / Helicity} }
  \myitem{double} { SPINUP(I) }
	{ cosine of the angle between the spin-vector of particle I
	  and the 3-momentum of the decaying particle, 
	specified in the lab frame } \\
        This scheme is neither general nor complete, but is chosen as
        the best compromise.
% since external legs to the partonic
%        subprocess will often be massless with a given chirality.
        The main foreseen application is $\tau$'s with a specific
	helicity. Typically a	
	relativistic $\tau^-$ ($\tau^+$) from a $W^-$ ($W^+$) has 
	helicity --1 (+1) (though this might be changed by the boost
	to the lab frame), so SPINUP(I)= --1 (+1). The use of a floating
	point number allows for the extension to the non-relativistic
	case. 
        Unknown or unpolarized particles should be given SPINUP(I)=9. The
        lab frame is the frame in which the four-vectors are
        specified.
	% 9 is used instead of 999 to allow for packing.        

\end{myitemize}

%
% Herwig uses separations at the femto scale (i.e. for b-decays the
%       separation in space WRT to the b-vertex is the thing of
%       importance, but since the b-vertex can be displaced at the 1mm 
%       level, machine precision (even for double) is not enough)
%
%
% NFUP, KFUP, Q2UP (which served this purpose in Pythia::PYUPPR) are not
% needed.

%\item Communication with Tauola/Photos is not addressed. (!)
%\item A generic file read/write routine could be supplied.
%\item Special treatment is normally needed for Deep Inelastic
%Scattering, wherein the recoil from the parton shower should not
%affect the incoming electron. The showering generators will be able to
%recognize these cases by the identity of the incoming particles.

%By allowing the incident quark to be both incoming and a daughter to
%the $\gamma^*$, the history will allow the showering event generator
%to know that the recoil from the parton shower should not alter 
%$\hat{s}$ and $\hat{t}$ of the interaction.

\begin{center}{\rule[1mm]{5in}{1mm}}\end{center} %%%%%line

\section*{Example: hadronic $t \bar{t} $ production }
\includegraphics[height=2.5in]{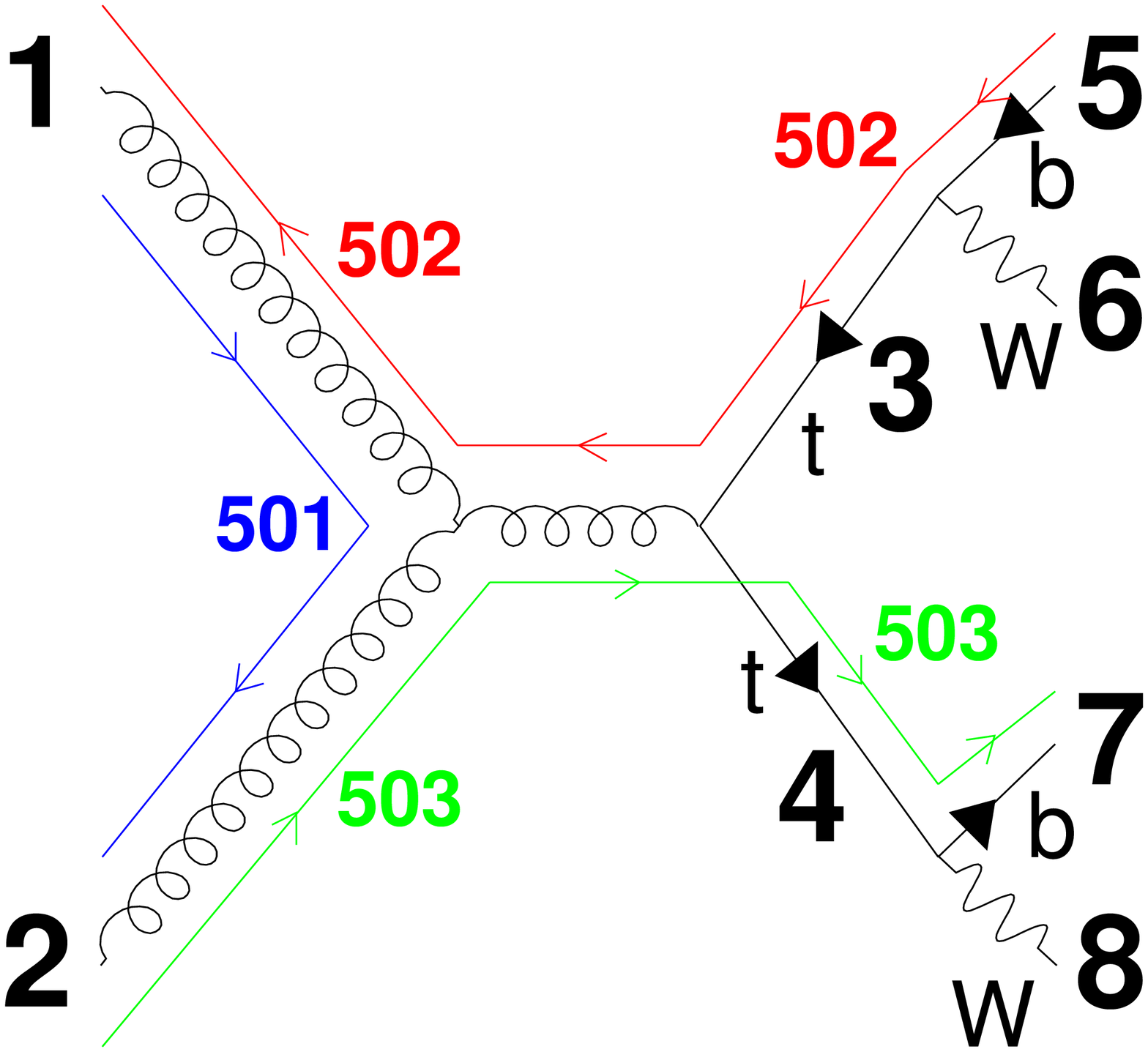}
\begin{center}
\begin{tabular}{ccccccc}
I & ISTUP(I) & IDUP(I)   & MOTHUP(1,I) & MOTHUP(2,I) 
                                        & ICOLUP(1,I) & ICOLUP(2,I) \\ \hline
1 &--1 &  21~($g$)               & 0   & 0         & 501 & 502     \\
2 &--1 &  21~($g$)               & 0   & 0         & 503 & 501     \\
%3 & +2 &  0\footnotesize{ (undefined)}
%                                 &  1 & 2        & 503 & 502     \\
3 & +2 & --6~($\bar{t}$)         &  1 & 2        &  0  & 502     \\
4 & +2 &  6~($t$)                &  1 & 2        & 503 &  0      \\
5 & +1 & --5~($\bar{b}$)         &  3 & 3        &  0  & 502     \\
6 & +1 & --24~($W^-$)            &  3 & 3        &  0  &  0      \\
7 & +1 &  5~($b$)                &  4 & 4        & 503 &  0      \\
8 & +1 &  24~($W^+$)             &  4 & 4        &  0  &  0      \\
\hline
\end{tabular}
\end{center}
The $t$ and $\bar{t}$ are given ISTUP=+2, which informs the \SHG\
to preserve their invariant masses when
showering and hadronizing the event. An intermediate s-channel gluon
has been drawn in the diagram, but since this graph cannot be usefully
distinguished from the one with a t-channel top exchange,
an entry has not been included for it in the event record.

The definition of a line as `color' or `anti-color' depends on the
orientation of the graph. This ambiguity is resolved  by defining 
color and anti-color according to the physical time order. 
A quark will always have its color tag ICOLUP(1,I) filled,
but never its anti-color tag ICOLUP(2,I). The reverse is true for an
anti-quark, and a gluon will always have information in both
ICOLUP(1,I) and ICOLUP(2,I) tags.

Note the difference in the treatment by the parton shower of the
above example, and an identical final state, where the intermediate
particles are not specified:
\begin{center}
\begin{tabular}{ccccccc}
I & ISTUP(I) & IDUP(I)   & MOTHUP(1,I) & MOTHUP(2,I) 
                                        & ICOLUP(1,I) & ICOLUP(2,I) \\ \hline
1 &--1        &  21~($g$)       &  0 & 0        & 501 & 502     \\
2 &--1        &  21~($g$)       &  0 & 0        & 503 & 501     \\
3 & +1        & --5~($\bar{b}$) &  1 & 2        &  0  & 502     \\
4 & +1        & --24~($W^-$)    &  1 & 2        &  0  &  0      \\
5 & +1        &  5~($b$)        &  1 & 2        & 503 &  0      \\
6 & +1        &  24~($W^+$)     &  1 & 2        &  0  &  0      \\
\hline
\end{tabular}
\end{center}
In this case the parton shower will evolve the $b,\bar{b}$ without
concern for the invariant mass of any pair of particles. 
Thus the parton shower may alter the invariant mass of the $Wb$
system (which may be undesirable if the $Wb$ was generated from a top
decay).

%The above example is ofcourse unphysical, in reality we only know that
%we started with an initial state $w^+e^-

\section*{Example: $gg \rightarrow gg$ }
\includegraphics[height=2.5in]{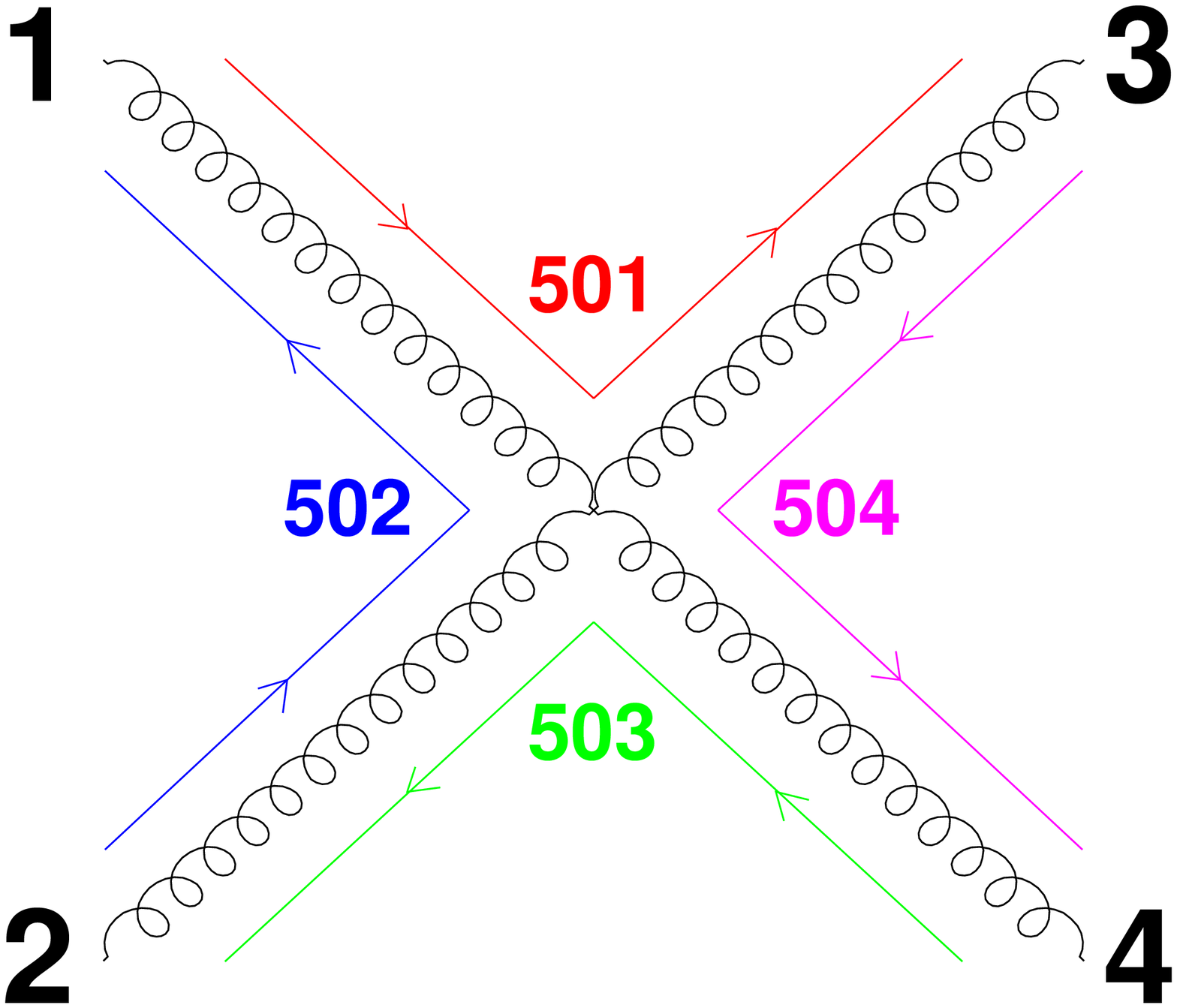}
\begin{center}
\begin{tabular}{ccccccc}
I & ISTUP(I) & IDUP(I) & MOTHUP(1,I) & MOTHUP(2,I) 
                                & ICOLUP(1,I) & ICOLUP(2,I) \\ \hline
1 &--1        &  21~($g$)    &  0 & 0        &   501 & 502   \\
2 &--1        &  21~($g$)    &  0 & 0        &   502 & 503   \\
3 & +1        &  21~($g$)    &  1 & 2        &   501 & 504   \\
4 & +1        &  21~($g$)    &  1 & 2        &   504 & 503   \\
\hline
\end{tabular}
\end{center}

\section*{Example: Baryon number violation in decays}
\includegraphics[height=2.5in]{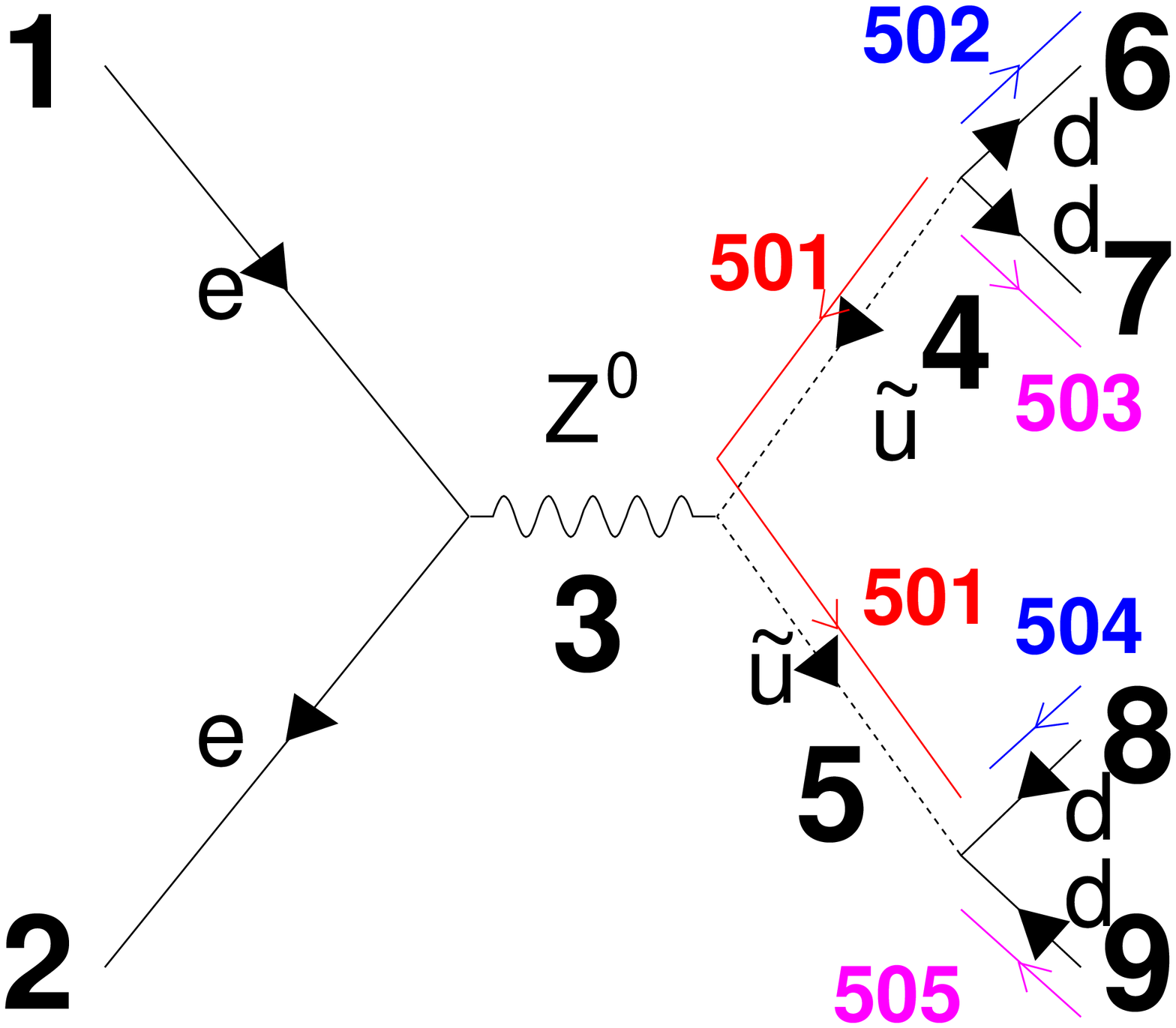}
\begin{center}
\begin{tabular}{ccccccc}
I & ISTUP(I) & IDUP(I)   & MOTHUP(1,I) & MOTHUP(2,I) 
                                        & ICOLUP(1,I) & ICOLUP(2,I) \\ \hline
1 &--1  &  11~($e^-$)                            &  0 & 0    &  0   &   0   \\
2 &--1  & --11~($e^+$)                           &  0 & 0    &  0   &   0   \\
3 & +2  &  23~($Z^0$)                            &  1 & 2    &  0   &   0   \\
4 & +2  &--1000002~($\stackrel{\sim}{\bar{u}}$)  &  3 & 3    &  0   &  501  \\
5 & +2  &  1000002~($\stackrel{\sim}{u}$)        &  3 & 3    & 501  &   0   \\
6 & +1  &  1~($d$)                               &  4 & 4    & 502  &   0   \\
7 & +1  &  1~($d$)                               &  4 & 4    & 503  &   0   \\
8 & +1  & --1~($\bar{d}$)                        &  5 & 5    &  0   &  504  \\
9 & +1  & --1~($\bar{d}$)                       &  5 & 5    &  0   &  505  \\
\hline
\end{tabular}
\end{center}
Three ``dangling'' color lines intersect at the vertex joining the
$\stackrel{\sim}{\bar{q}},q,q'$ 
(and $\stackrel{\sim}{q},\bar{q},\bar{q}'$), which
corresponds to a Baryon number source (sink) of +1 (-1), and will be
recognizable to the \SHG s.

\section*{Example: Baryon number violation in production}
\includegraphics[height=2.5in]{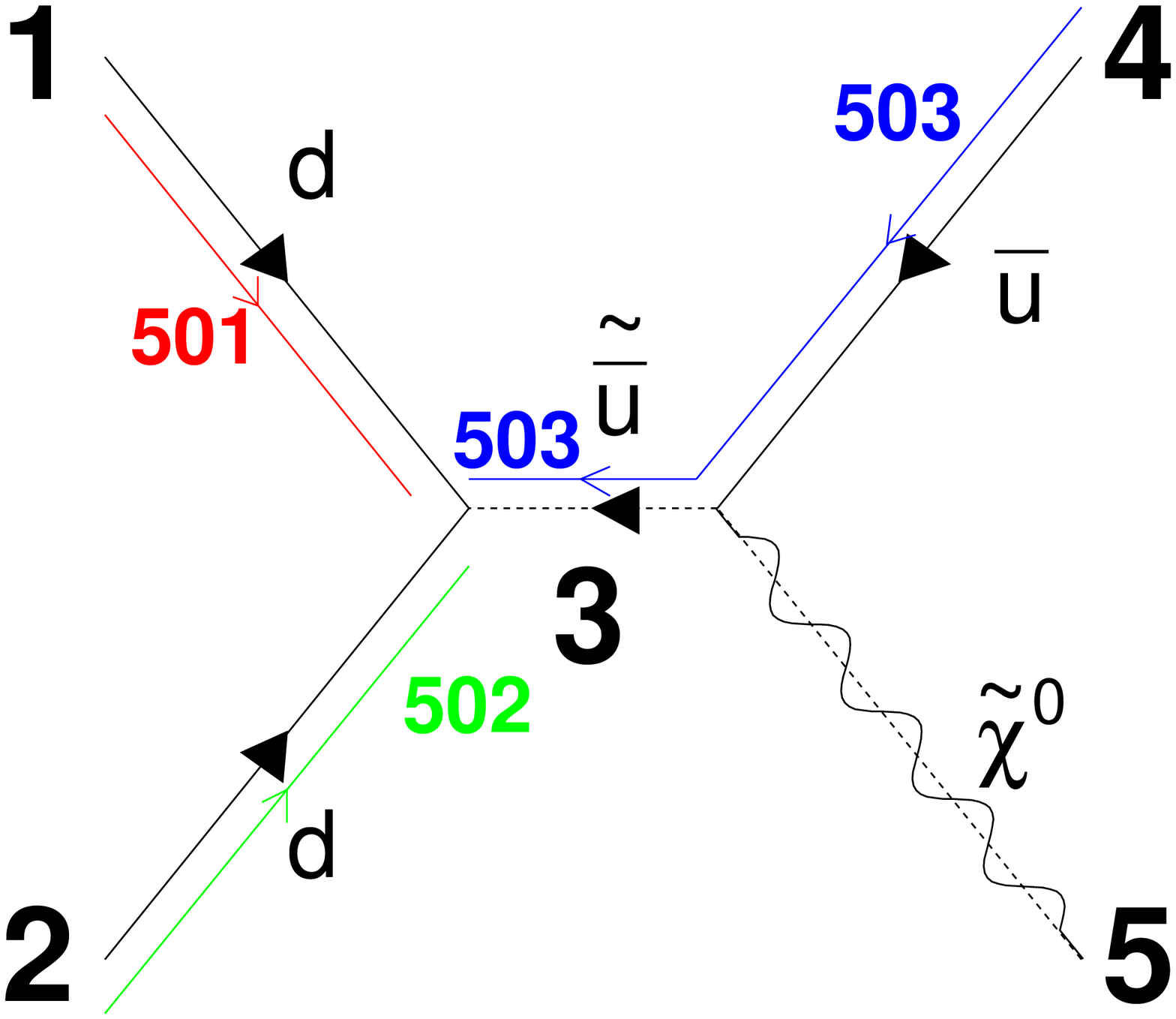}
\begin{center}
\begin{tabular}{ccccccc}
I & ISTUP(I) & IDUP(I)               & MOTHUP(1,I) & MOTHUP(2,I) 
                                        & ICOLUP(1,I) & ICOLUP(2,I) \\ \hline
1 &--1 &  1~($d$)                               & 0 & 0   & 501  & 0        \\
2 &--1 &  1~($d$)                               & 0 & 0   & 502  & 0        \\
3 & +2 & --1000002~($\stackrel{\sim}{\bar{u}}$) & 1 & 2   &  0   & 503      \\
4 & +1 & --2~($\bar{u}$)                        & 3 & 3   &  0   & 503      \\
5 & +1 &  1000022~($\stackrel{\sim}{\chi}^0$)   & 3 & 3   &  0   & 0        \\
\hline
\end{tabular}
\end{center}

\section*{Example: deep inelastic scattering }
\label{example_dis.eps}
\includegraphics[height=2.5in]{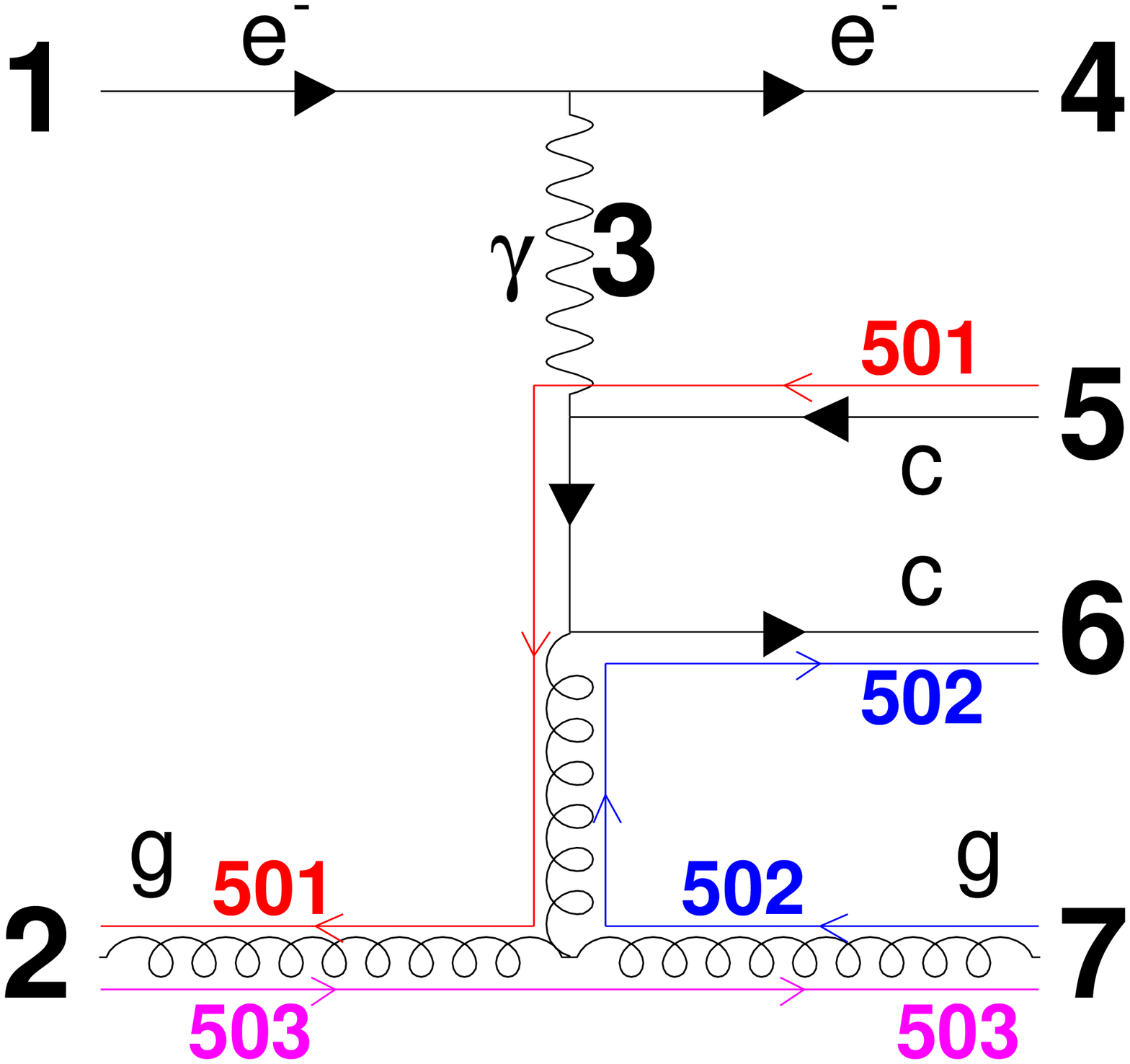}
\begin{center}
\begin{tabular}{ccccccc}
I & ISTUP(I) & IDUP(I) & MOTHUP(1,I) & MOTHUP(2,I) 
                                & ICOLUP(1,I) & ICOLUP(2,I) \\ \hline
1 &--1        &  11~($e^-$)    &  0 &  0       &   0   &  0    \\
2 &--1        &  21~($g$)      &  0 &  0       &  503  & 501   \\
3 &--2        &  22~($\gamma$) &  1 &  0       &   0   &  0    \\
4 & +1        &  11~($e^-$)    &  1 &  0       &   0   &  0    \\
5 & +1        &  -4~($\bar{c}$)&  2 &  3       &   0   & 501   \\
6 & +1        &  4~($c$)       &  2 &  3       &  502  &  0    \\
7 & +1        &  21~($g$)      &  2 &  3       &  503  & 502   \\
\hline
\end{tabular}
\end{center}

For DIS, the $x$ and $q^2$ of the $\gamma$ should not be altered by
the parton shower, so the $\gamma$ is given ISTUP=--2. We have not
specified the internal quark and gluon lines which will be dressed by
the parton shower, such that the partonic event configuration may be
drawn as follows, \\
\includegraphics[height=2.5in]{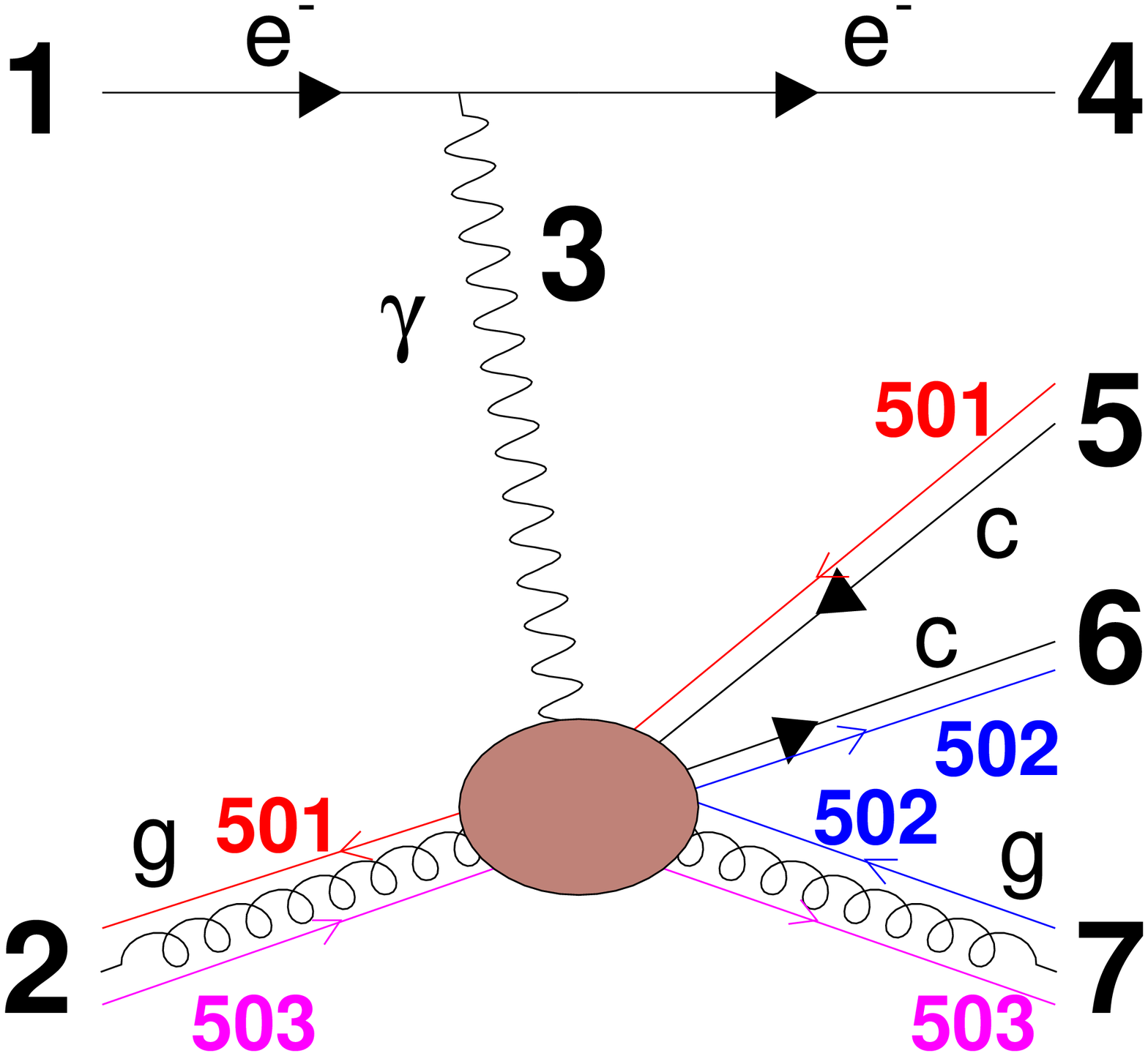}. \\
If information about the quark and gluon propagators is desired (i.e\
for human readability), then those entries may be included with status
code ISTUP=+3.


\begin{thebibliography}{99}



%\cite{Pukhov:1999gg}
\bibitem{Pukhov:1999gg}
A.~Pukhov {\it et al.},
%``CompHEP: A package for evaluation of Feynman diagrams and
%integration over multi-particle phase space. User's manual for version 33,''
hep-ph/9908288.
%%CITATION = HEP-PH 9908288;%%

%\cite{Ishikawa:1993qr}
\bibitem{Ishikawa:1993qr}
T.~Ishikawa, T.~Kaneko, K.~Kato, S.~Kawabata, Y.~Shimizu and H.~Tanaka
                  [MINAMI-TATEYA group Collaboration],
%``GRACE manual: Automatic generation of tree amplitudes in
% Standard Models: Version 1.0,''
KEK-92-19.

%\cite{Stelzer:1994ta}
\bibitem{Stelzer:1994ta}
T.~Stelzer and W.~F.~Long,
%``Automatic generation of tree level helicity amplitudes,''
Comput.\ Phys.\ Commun.\  {\bf 81}, 357 (1994)
[hep-ph/9401258].
%%CITATION = HEP-PH 9401258;%%

%\cite{Berends:1991ax}
\bibitem{Berends:1991ax}
F.~A.~Berends, H.~Kuijf, B.~Tausk and W.~T.~Giele,
%``On the production of a W and jets at hadron colliders,''
Nucl.\ Phys.\ B {\bf 357}, 32 (1991).
%%CITATION = NUPHA,B357,32;%%

%  %\cite{Katsanevas:1998fb}
%  \bibitem{Katsanevas:1998fb}
%  S.~Katsanevas and P.~Morawitz,
%  %``SUSYGEN-2.2: A Monte Carlo event generator for MSSM sparticle    
%  % production at e+ e- colliders,''
%  Comput.\ Phys.\ Commun.\  {\bf 112}, 227 (1998)
%  [hep-ph/9711417].
%  %%CITATION = HEP-PH 9711417;%%

%\cite{Caravaglios:1999yr}
\bibitem{Caravaglios:1999yr}
F.~Caravaglios, M.~L.~Mangano, M.~Moretti and R.~Pittau,
%``A new approach to multi-jet calculations in hadron collisions,''
Nucl.\ Phys.\ B {\bf 539}, 215 (1999)
[hep-ph/9807570].
%%CITATION = HEP-PH 9807570;%%


%\cite{Corcella:2001bw}
\bibitem{Corcella:2001bw}
G.~Corcella {\it et al.},
%``HERWIG 6: An event generator for hadron emission reactions with  interfering gluons (including supersymmetric processes),''
JHEP {\bf 0101}, 010 (2001)
[hep-ph/0011363].
%%CITATION = HEP-PH 0011363;%%

%                                     Isajet
%\cite{Baer:1999sp}
\bibitem{Baer:1999sp}
H.~Baer, F.~E.~Paige, S.~D.~Protopopescu and X.~Tata,
%``ISAJET 7.48: A Monte Carlo event generator for p p, anti-p p, and  
% e+ e- reactions,''
hep-ph/0001086.
%%CITATION = HEP-PH 0001086;%%

%\cite{Sjostrand:2001wi}
\bibitem{Sjostrand:2001wi}
T.~Sjostrand, P.~Eden, C.~Friberg, L.~Lonnblad, G.~Miu, S.~Mrenna and E.~Norrbin,
%``High-energy-physics event generation with PYTHIA 6.1,''
Comput.\ Phys.\ Commun.\  {\bf 135}, 238 (2001)
[hep-ph/0010017].
%%CITATION = HEP-PH 0010017;%%

%\cite{Mangano:2001xp}
\bibitem{Mangano:2001xp}
M.~L.~Mangano, M.~Moretti and R.~Pittau,
hep-ph/0108069.
%%CITATION = HEP-PH 0108069;%%

%\cite{Belyaev:2000wn}
\bibitem{Belyaev:2000wn}
A.~S.~Belyaev {\it et al.},
%``CompHEP-PYTHIA interface: Integrated package for the collision events  generation based on exact matrix elements,''
hep-ph/0101232.
%%CITATION = HEP-PH 0101232;%%

\bibitem{Sjostrand:2001doc}
T. Sj\"ostrand, L. L\"onnblad and S. Mrenna, LU~TP~01-21, 
[hep-ph/0108264].

\bibitem{HEPEVT}  T.~Sj\"{o}strand et al., in ``Z physics at LEP 1'',
  eds. G.~Altarelli, R.~Kleiss and C.~Verzegnassi, 
	CERN 89-08 (Geneva, 1989), Vol. 3, p. 327.

%\cite{Dobbs:2001ck}
\bibitem{Dobbs:2001ck}
M.~Dobbs and J.~B.~Hansen,
%``The HepMC C++ Monte Carlo event record for High Energy Physics,''
Comput.\ Phys.\ Commun.\  {\bf 134}, 41 (2001).
%%CITATION = CPHCB,134,41;%%

%\cite{Garren:2000st}
\bibitem{Garren:2000st}
L.~Garren, I.~G.~Knowles, T.~Sjostrand and T.~Trippe,
%``Monte Carlo particle numbering scheme: in Review of Particle Physics (RPP 2000),''
Eur.\ Phys.\ J.\ C {\bf 15}, 205 (2000).
%%CITATION = EPHJA,C15,205;%%

%\cite{Plothow-Besch:1993qj}
\bibitem{Plothow-Besch:1993qj}
H.~Plothow-Besch,
%``PDFLIB: A Library of all available parton density functions of the
%nucleon, the pion and the photon and the corresponding alpha-s calculations,''
Comput.\ Phys.\ Commun.\  {\bf 75}, 396 (1993);
CERN Program Library Long Writeup W5051 (2000);
refer to  {\tt http://consult.cern.ch/writeup/pdflib/}.
%%CITATION = CPHCB,75,396;%%

\end{thebibliography}
\end{document}